\documentclass{ws-ijmpcs}
\usepackage{amssymb,amsmath,amsfonts,latexsym,graphicx,epsfig}
\usepackage{color} 

\newcommand{\bea}{\begin{eqnarray}}
\newcommand{\eea}{\end{eqnarray}}
\newcommand{\be}{\begin{equation}}
\newcommand{\ee}{\end{equation}}
\newcommand{\nn}{\nonumber}
\newcommand{\nnnl}{\nonumber\\}

\newcommand{\msbar}{$\overline{\rm MS}\,\,$}

\newcommand{\chpt}{ChPT\,\,}

\begin{document}

\markboth{M.A. Ivanov}
{Pion polarizabilities: Theory vs Experiment}

%
\catchline{}{}{}{}{}
%

\title{Pion polarizabilities: Theory vs Experiment}

\author{M.A. Ivanov}

\address{Bogoliubov Laboratory of Theoretical Physics,\\
Joint Institute for Nuclear Research, \\
141980 Dubna (Moscow region), Russia \\
ivanovm@theor.jinr.ru}

\maketitle


\begin{abstract}

The values of charged pion polarizabilities obtained 
in the framework of chiral perturbation theory at the level
of two-loop accuracy are compared with the experimental result 
recently reported by COMPASS Collaboration.
It is found that the calculated value for the dipole polarizabilities   
$(\alpha-\beta)_{\pi^\pm}= (5.7\pm 1.0)\times 10^{-4}\,{\rm fm}^3$
fits quite well the experimental result 
$(\alpha-\beta)_{\pi^\pm} =  
(4.0  \pm 1.2_{\rm stat} \pm 1.4_{\rm syst}) \times 10^{-4}\,{\rm fm}^3$.


\end{abstract}

\ccode{PACS numbers:11.30.Rd;12.38.Aw;12.39.Fe;}

\section{Introduction}

Almost ten years ago we have evaluated the amplitude
for $\gamma\gamma\rightarrow\pi\pi$  
in the framework of chiral perturbation theory (\chpt) 
\cite{Weinberg:1978kz,GLann,GLnpb} at two-loop order.
It was done for neutral pions \cite{Gasser:2005ud}
as well as for charged piones \cite{Gasser:2006qa}.
The obtained results were found to be in agreement with the only previous 
calculation performed at this accuracy \cite{BGS,Burgi},
provided that the same set of low-energy constants (LECs) is used.
With updated LECs at order $p^4$ \cite{CGLpipi,BijnensTalavera}, 
we found \cite{Gasser:2006qa} for the dipole polarizabilities the values
\begin{eqnarray}
(\alpha-\beta)_{\pi^\pm} &=& (5.7\pm1.0)\times 10^{-4}\, {\rm fm}^3,
\nonumber\\
(\alpha+\beta)_{\pi^\pm} &=& 0.16\times 10^{-4}\, {\rm fm}^3.
\label{eq:GIS}
\end{eqnarray}
At that time the MAMI Collaboration \cite{MAMI} has  
reported the experimental result
\bea
\label{eq:polarMAMI}
(\alpha-\beta)_{\pi^\pm}= (11.6\pm 1.5_{\rm stat}
\pm 3.0_{\rm syst}\pm 0.5_{\rm mod})\times 10^{-4}\,{\rm fm}^3\,.
\eea
The index ``mod'' denotes
the uncertainty generated by the theoretical models used to analyze
the data. The ChPT calculation was clearly in conflict with the MAMI
result, see also \cite{schererreview} for a recent discussion. 

The COMPASS collaboration at CERN has recently investigated pion Compton 
scattering by using the Primakoff effect.
The pion polarizability has been  determined to be \cite{Adolph:2014kgj} 
\bea
\label{eq:polarCOMPASS}
(\alpha-\beta)_{\pi^\pm} =  
(4.0  \pm 1.2_{\rm stat} \pm 1.4_{\rm syst}) \times 10^{-4}\,{\rm fm}^3
\eea
under  the assumption $(\alpha+\beta)_{\pi^\pm}=0$.
This result is in agreement with the expectation from 
chiral perturbation theory. 

The concept of the polarizability of molecules, atoms and nuclei
was applied for the first time to hadrons in 
Refs.~\cite{Klein:1955zz,Baldin:1960,Petrunkin:1961}.
By using the general properties of quantum field theory
it was shown that an expansion of the Compton scattering amplitude
for hadrons with spin one half in small photon energy up to the second
order contains two structure parameters called the electric and magnetic
hadron polarizabilities.  The classical sum rule for these quantities
has been derived in Ref.~\cite{Petrunkin:1964}.
Further theoretical investigation of the pion polarizabilities
has been pursued since the early 1970s. 
In the current algebra + PCAC approach of Terent'ev~\cite{Terentev}, 
the fundamental low-energy theorem has been proven which allows one to 
relate the pion polarizability to the ratio $\gamma=h_A(0)/h_V(0)$ 
of the vector and axial form factors in radiative pion decay
$\pi\to e\nu\gamma$. By using recent precise
measurements of the pion weak form factors by the PIBETA collaboration
 \cite{Bychkov:2008ws} one finds 
$\alpha_{\pi^\pm}=-\beta_{\pi^\pm}=2.78(10)\times 10^{-4}\, {\rm fm}^3$.  

There were many calculations of the pion polarizabilities
by employing various models: the linear $\sigma$-model with 
quarks~\cite{Lvov:1980st}, the chiral quark model~\cite{Volkov:1980cj},
the superconductor quark model~\cite{Volkov:1985nk},
some chiral models~\cite{Bernard:1988gp} and so on.

Almost all of them except Terent'ev approach predicted a value
of the electric polarizability within the range 
\begin{equation}
     4.0\times 10^{-4}\, {\rm fm}^3 \le\alpha_{\pi^\pm} 
 \le 6.0 \times 10^{-4}\, {\rm fm}^3
\label{eq:LVR}
\end{equation}
which we call large-valued results.
We note that models not based on a chiral Lagrangian, i.e.,
dispersion relations and finite-energy sum rules, also obtained the
polarizability within this range of values~\cite{Filkov:1982cx,Fil'kov:2005ss}.
The pion and kaon polarizabilities have been calculated in 
the quark confinement model~\cite{Ivanov:1991kw} in which the emphasis
is placed on quark confinement and the composite nature of hadrons.
It was found for charged pions 
$\alpha_{\pi^\pm}\sim 3.6\times 10^{-4}\, {\rm fm}^3$ which is smaller
than the large-valued results but slightly larger than Terent'ev's prediction.  

The first correct calculation of the  cross section $\gamma\gamma\to\pi\pi$
within chiral perturbation theory  to next-to-leading order (one-loop
accuracy) was performed in \cite{Bijnens:1987dc}. It was shown in 
\cite{Donoghue:1989sq} that chiral symmetry relates the low-energy
constants (LECs) appearing in the $\gamma\gamma\to\pi\pi$-amplitude
with the axial form factor $h_A(0)$. Thus it was shown explicitly that
Terent'ev's low-energy theorem follows from one-loop calculation
of the  $\gamma\gamma\to\pi\pi$ process within chiral perturbation theory.

Note that the  axial form factor $h_A(0)$ can be expressed through the
dispersion integral of the difference of the vector and axial spectral 
densities~\cite{Das:1967ek}. By using this sum rule the pion
polarizability was estimated in \cite{Dorokhov:2003kf} and found to be
in perfect agreement with chiral perturbation theory.

An actual two-loop ChPT calculation of the   $\gamma\gamma\to\pi\pi$ 
amplitude was done in \cite{BGS} (neutral pions) and
\cite{Burgi} (charged pions). Because the effective Lagrangian at order $p^6$ 
was not available at that time, the ultraviolet divergences were 
evaluated in the \msbar scheme,  then dropped and replaced with 
a corresponding polynomial in the external momenta. 
The three new counterterms which enter at this order in 
the low-energy expansion were estimated with resonance saturation. 
Whereas such a procedure is  legitimate from a technical point of view, 
it does not make  use of the full information
provided by chiral symmetry.

Later on, considerable progress has been made in this field,
 both in theory and experiment. As for theory, 
the Lagrangian at order $p^6$
 has been constructed \cite{SchererFearing,BCE1}, and its divergence 
structure has been determined \cite{BCE2}. 
This provides an important check on the above calculations: adding the
 counterterm contributions from the $p^6$ Lagrangian to the  \msbar 
amplitude evaluated in \cite{BGS} and in 
\cite{Burgi} must provide a scale independent result.
Also in the  theory, improved techniques to evaluate the two-loop diagrams 
that occur in these amplitudes have been developed \cite{GS}.
The updated calculation of the   $\gamma\gamma\to\pi\pi$ amplitude to two loops
was then performed in   \cite{Gasser:2005ud} (neutral pions)
and  \cite{Gasser:2006qa} (charged pions). The final results
for the pion polarizabities were presented in a rather compact algebraic
form. By using updated values for the LECs one obtains the values
of the pion polarizabilities given in Eq.~(\ref{eq:GIS}).

A comprehensive review of the modern status of this field maybe found
in Refs.~\cite{Petrunkin:1981me,Holstein:2013kia}.

Finally, one has to mention that research on pion polarizabities 
using lattice simulation is currently conducted by several groups, see, 
for instance, 
Refs.~\cite{Luschevskaya:2014lga,Lujan:2014kia,Detmold:2009dx,Hu:2007ts}.

\section{Definition of pion polarizabilities} 
\label{sec:definition}
\begin{figure}[ht]
\centerline{\includegraphics[width=10cm]{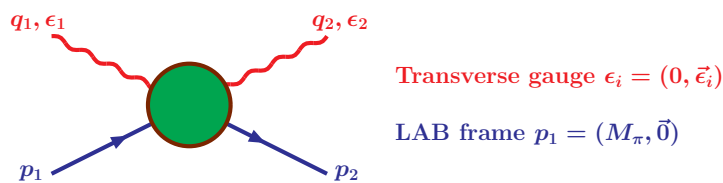}}
\vspace*{8pt}
\caption{A schematic illustration of Compton scattering by a pion.
\label{fig:Compton}}
\end{figure}

The electric ($\alpha_H$) and magnetic  ($\beta_H$) polarizabilities
characterize the response of hadron to two-photon interactions.
These quantities must be considered as fundamental as the electromagnetic 
mean square radii, static magnetic moments, etc. They are defined 
by the expansion of the Compton scattering amplitude in small
photon momenta and energies. Since our interest here is the pion
polarizabilities, we plot in Fig.~\ref{fig:Compton} the diagram describing
the  Compton scattering by charged pion.
Expanding the  Compton scattering amplitude in small photon momenta and 
energies, one finds
\begin{eqnarray}
T_{\gamma\pi^+\to\gamma\pi^+} &=& 
\underbrace{
-2\,e^2\,\vec\epsilon_1\cdot\vec\epsilon_2}_
{\mbox{\small\bf Born term}}
\nonumber\\[1.2ex]
&+&8\,\pi\,M_\pi
\underbrace{
 \Big\{ \alpha_\pi\,\omega_1\omega_2\,
\vec\epsilon_1\cdot\vec\epsilon_2
       + \beta_\pi \,(\vec\epsilon_1\times \vec q_1)\cdot
 (\vec\epsilon_2\times \vec q_2)\Big\}}_
{\mbox{\small\bf el-mag  polarizabilities}}
+\ldots
\label{eq:def}
\end{eqnarray}
It is convenient to use the linear combinations of the electric
and magnetic polarizabilities: $(\alpha-\beta)_\pi$ and $(\alpha+\beta)_\pi$
which are obtained from the helicity flip and helicity non-flip amplitudes,
respectively. 

As follows from the definition, the dipole pion polarizabilities
are proportional to 
$ 
\alpha_\pi(\beta_\pi)\sim \frac{\alpha}{M_\pi}\frac{1}{\Lambda^2}
\approx 4\times 10^{-4}\,{\rm fm}^3
$
where the  hadronic scale $\Lambda\sim 4\pi F_\pi\sim 1\,{\rm GeV}$ 
was used.  Then a natural choice of units for the polarizabilities is
$10^{-4}\,{\rm fm}^3$.

\section{Effective Lagrangian} 
\label{sec:effective}

We consider  an effective Lagrangian of QCD with two flavors in the
isospin symmetry limit $m_u=m_d=\hat m$.
At  next-to-next-to-leading order (NNLO), one has \cite{GLann}
\bea
{\mathcal L}_{\rm eff}={\mathcal L}_2+{\mathcal L}_4 +{\mathcal L}_6\,.
\eea
The subscripts refer to the chiral order. 
The expression for ${\mathcal L}_2$ is
\bea
\label{eq:l2}
{\mathcal L}_{\, 2} &=&\frac{F^2}{4}\langle D_\mu U\,D^\mu U^\dagger
        +M^2(U + U^\dagger)\rangle\, ,\nnnl
D_\mu U &=& \partial_\mu U -i(QU-UQ)A_\mu\, , \, \
Q=\frac{e}{2}{\rm diag}(1,-1)\, ,
\eea
where $e$ is the electric charge, and $A_\mu$ denotes the electromagnetic 
field.
The quantity $F$ denotes the pion decay constant 
in the chiral limit,  and $M^2$ is the leading term in the quark mass
expansion of the pion (mass)$^2$, $M_\pi^2=M^2(1+O(\hat m))$.
Further, the brackets $\langle\ldots\rangle$ denote a trace in flavor space.
In Eq. (\ref{eq:l2}), we have retained only the terms relevant for 
the present application,
i.e., we have dropped additional external fields.
We choose the unitary $2\times 2$ matrix $U$ in the form
\bea
U &=& \sigma + i\, \pi/F\,, 
\hspace{.3cm} \sigma^2 + \frac{\pi^2}{F^2} = {\mathbf 1}_{2\times 2}\,,
\hspace{.3cm}\pi=\left( \mbox{$\begin{array}{cc} \pi^0 & \sqrt{2}\, \pi^+
         \\ \sqrt{2}\, \pi^- &- \pi^0 \end{array}$} \right)  \; .
\eea
The  Lagrangian at NLO  has the structure~\cite{GLann}
\bea
{\mathcal L}_4=\sum_{i=1}^{7} l_iK_i+\sum_{i=1}^{3}h_i\bar K_i=\frac{l_1}{4}\langle
D_\mu U\,D^\mu U^\dagger\rangle^2
+\cdots\, ,
\eea
where  $l_i,h_i$ denote low-energy couplings, 
not fixed by chiral symmetry.  
At NNLO, one has \cite{BCE1,BCE2,SchererFearing}
\bea
{\mathcal L}_6=\sum_{i=1}^{57} c_i P_i\,.
\eea
As was shown in Ref.~\cite{Haefeli:2007ty}
the number of operators $P_i$ can be reduced by at least one from 57 to 56. 
For the explicit expressions of the polynomials $K_i,\bar K_i$ and $P_i$, 
we refer the reader to
Refs.~\cite{GLann,BCE1,BCE2,SchererFearing}. The vertices relevant for
$\gamma\gamma\rightarrow \pi^+\pi^-$ involve  $l_1, \ldots, l_6$ 
{}from ${\mathcal L}_4$ and several $c_i$'s 
{}from ${\mathcal L}_6$, see below.

The couplings $l_i$ and $c_i$ absorb the divergences at order~$p^4$ and 
$p^6$, respectively,
\bea\label{eq:lici}
l_i &=& (\mu\,c)^{d-4}
\left\{l_i^r(\mu,d) + \gamma_i\,\Lambda\right\}\,, 
\nn\\[2mm]
c_i&=&\frac{(\mu\,c)^{2(d-4)}}{F^2}
\left\{
c_i^r(\mu,d) - \gamma_i^{(2)}\,\Lambda^2
       -(\gamma_i^{(1)}+\gamma_i^{(L)}(\mu,d))\,\Lambda\right\}\,, 
\nn\\[2mm]
\Lambda &=& \frac{1}{16\,\pi^2 (d-4)}\,,
\, \ln c = -\frac{1}{2}\left\{\ln 4\pi +\Gamma'(1)+1\right\}\,.
\eea
The physical couplings are $l_i^r(\mu,4)$ and $c_i^r(\mu,4)$, denoted by
$l_i^r,c_i^r$ in the following.
The coefficients $\gamma_i$ are given in \cite{GLann}, 
and $\gamma_i^{(1,2,L)}$ are tabulated in \cite{BCE2}.
We shall use the scale independent quantities 
$\bar l_i$ introduced in \cite{GLann}, 
\begin{eqnarray} 
\label{RGE}
l_i^r &=& \frac{\gamma_i}{32\pi^2}\,({\bar l}_i + l)\,,
\end{eqnarray}
where the {\it chiral logarithm} is $l=\ln(M^2_\pi/\mu^2)$. We shall 
use \cite{CGLpipi}
\bea 
\label{eq:LECsp4}
\bar l_1=-0.4\pm 0.6\,,\,\,\bar l_2=4.3\pm0.1\,,\,\,
\bar l_3=2.9\pm2.4\,,\,\,
\bar l_4=4.4\pm0.2\,,
\eea
and 
\bea
\label{eq:LECs65}
\bar l_\Delta\doteq \bar l_6-\bar l_5=3.0\pm 0.3
\eea
obtained from radiative pion decay to two loop
accuracy \cite{BijnensTalavera,Geng}.

The constants $c_i^r$ occur in the combinations
 \begin{eqnarray}
a_1^r &=& -4096\pi^4\left( 6\, c_{6}^r+ c_{29}^r -c_{30}^r -3\, c_{34}^r 
          + c_{35}^r + 2\,c_{46}^r - 4\, c_{47}^r + c_{50}^r \right)\,,
\nn\\
&&\nn\\
a_2^r &=& 256\pi^4\left( 
8\,c_{29}^r - 8\,c_{30}^r + c_{31}^r + c_{32}^r- 2\,c_{33}^r 
+4\, c_{44}^r +8\, c_{50}^r -4\, c_{51}^r \right)\,,
\nn \\
&&\nn\\
b^r &=&-128\pi^4
     \left(c_{31}^r + c_{32}^r - 2\,c_{33}^r -4\, c_{44}^r \right)\,.\nonumber
\end{eqnarray}
As follows from the resonance exchange model ~\cite{Burgi}
\bea\label{eq:LECsp6}
\Big(a_1^r, a_2^r, b^r\Big)=
\Big(-3.2,0.7,0.4\Big)\qquad  [{\mbox{ present work }}]\,.
\eea
The values of these constants were obtained in the 
ENJL model \cite{BijnensPrades} $(a_1^r, a_2^r, b^r)=(-8.7,5.9,0.38)$
One can see that only $b^r$ agrees in the two approaches.
We shall use $b^r=0.4\pm 0.4$.
The combinations $(\alpha \pm\beta)_{\pi^\pm}$ are independent of $a_2^r$ 
and are determined precisely by the chiral expansion to two loops,
once $a_1^r$ is fixed. We will then simply display this quantity as
a function of $a_1^r$ - the result turns out to be rather independent of its 
exact value.

\section{Evaluation of the diagrams}
\label{sec:diagrams}

\begin{figure}[pb]
\begin{center}
\centerline{\includegraphics[width=10cm]{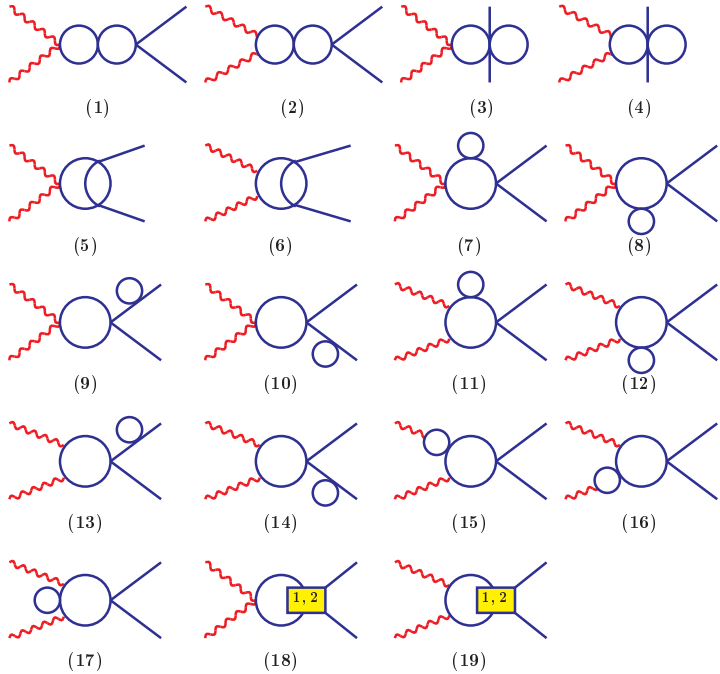}}
\end{center}
\caption{\small
A set of two-loop diagrams generated by~${\mathcal L}_2$
and one-loop diagrams generated by~${{\mathcal L}_4.}$} 
\label{fig:two-loop}
\end{figure}

\begin{figure}[bp]
\begin{center}
\centerline{\includegraphics[width=8cm]{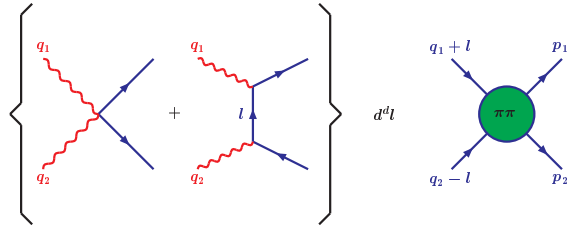}}
\end{center}
\caption{\small
Construction scheme for the diagrams in Fig.~\ref{fig:two-loop}.} 
\label{fig:scheme}
\end{figure}

\begin{figure}[bp]
\begin{center}
\epsfig{figure=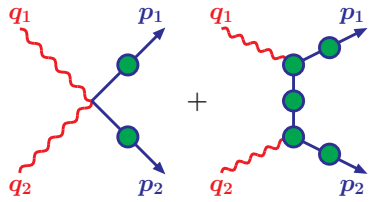,width=5cm} 
\end{center}
\caption{\small A class of one-particle  reducible diagrams. 
The filled in circles summarize self-energy and vertex corrections.} 
\label{fig:reducible}
\end{figure}

\begin{figure}[bp]
\begin{center}
\centerline{\includegraphics[width=6cm]{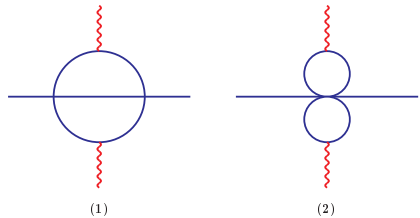}}
\end{center}
\caption{\small Acnode and butterfly diagrams.}
\label{fig:acnode}
\end{figure}

The lowest-order contributions to the scattering amplitude 
are described by  tree- and one-loop  diagrams. 
These contributions were calculated in \cite{Bijnens:1987dc}.
The two-loop diagrams are displayed in Figs.~\ref{fig:two-loop}, 
\ref{fig:reducible} and \ref{fig:acnode}.
The  two-loop diagrams in Fig.~\ref{fig:two-loop} may be generated 
according to the scheme indicated in 
Fig.~\ref{fig:scheme}, where the filled in blob denotes
the $d$-dimensional elastic $\pi\pi$-scattering amplitude   
at one-loop accuracy, with two pions off-shell. 

The diagrams shown in Fig.~\ref{fig:reducible} 
may be reduced to tree-diagrams by using Ward
identities. They sum up to the expression
\bea
&&
2\, Z_\pi\, g^{\mu\nu} 
- \left\{(2 p_1-q_1)^\mu (2 p_2-q_2)^\nu 
         \left[\frac{1}{M^2_\pi-t}-Z_\pi R(t)\right] 
+ {\rm crossed } \right\},
\eea
where $Z_\pi$ is the pion renormalization constant. The function
$R(t)$ starts at order $1/F^4_\pi$ and can be obtained from
the full pion propagator.

\begin{figure}[ht]
\begin{center}
\epsfig{figure=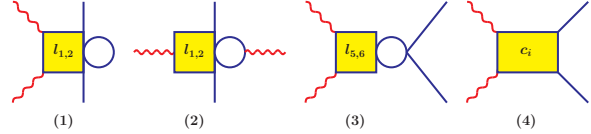,width=8cm}  
\end{center}
\caption{\small The remaining diagrams at order~$p^6$: 
one-loop graphs generated by ${\mathcal L}_4$, and counterterm 
contributions  from ${\mathcal L}_6$.} 
\label{fig:lecs}
\end{figure}

Two further diagrams are displayed in Fig.~\ref{fig:acnode}.
The first one - called  ``acnode'' in the literature - may again be evaluated
by use of a dispersion relation, see \cite{Gasser:2005ud} .
The second one is trivial to evaluate, because it is a product of one-loop
diagrams. 
The~remaining diagrams at order~$p^6$ are shown in Fig.~\ref{fig:lecs}.

The evaluation of the diagrams was done in the manner
described in \cite{Gasser:2005ud,GS} by invoking  FORM \cite{Vermaseren}.
In particular, we have verified that the
counterterms from the Lagrangian ${\mathcal L}_6$ \cite{BCE2} 
remove all ultraviolet divergences, which is a very non-trivial 
check on our calculation. Furthermore, 
we have checked that the (ultra-violet finite) amplitude so obtained 
is scale independent.

\section{Chiral expansion for pion polarizabilities}
Using the same notation as in \cite{Burgi}, we find for the 
{\it dipole} polarizabilities
\begin{equation}
\label{eq:alphabeta}
(\alpha \pm \beta)_{\pi^+} = 
           \frac{\alpha}{16\,\pi^2\,F_\pi^2\,M_\pi}\,
\left\{c_{1\pm} + \frac{M_\pi^2\,d_{1\pm}}{16\,\pi^2\,F_\pi^2} +O(M^4_\pi)
\right\}\,,
\end{equation}
where 
\begin{eqnarray}
c_{1+} &=& 0, \qquad c_{1-}=\frac23\,\bar l_\Delta\, ,
\nnnl
d_{1+} &=& 8\,b^r 
-\frac{4}{9}\left\{ l\,\left( l+\frac{1}{2}\,\bar{l}_1
                   +\frac{3}{2}\,\bar{l}_2\right)
-\frac{53}{24}\,l  + \frac{1}{2}\,\bar{l}_1
+\frac{3}{2}\,\bar{l}_2+\frac{91}{72} +\Delta_+ \right\} \;\; , 
\nnnl
d_{1-} & = & a_1^r +8\,b^r  -\frac{4}{3}
  \left\{l\left(\bar{l}_1-\bar{l}_2+\bar{l}_{\Delta}-\frac{65}{12}\right)
 - \frac{1}{3}\,\bar{l}_1 -\frac{1}{3}\,\bar{l}_2+\frac{1}{4}\,\bar{l}_3
 -\bar{l}_{\Delta}\bar{l}_4
\right. 
\nnnl
&&
\left.
+\frac{187}{108} + \Delta_-\right\}\,,
\label{polchi3}
\eea
with
\bea
\Delta_+ & = & \frac{8105}{576} - \frac{135}{64}\,\pi^2\,, \qquad
\Delta_-   =   \frac{41}{432} - \frac{53}{64}\,\pi^2\,. 
\label{polchi4}
\eea
It would be interesting to numerically compare the values of $\Delta_\pm$ given
by Eq.~(\ref{polchi4}) with those obtained in Refs.~\cite{Burgi}.
One has
\[
 \Delta_+  =  \left\{\begin{array}{l}
          -6.75 \qquad\text{our} \\
          -8.69 \qquad\text{Burgi}
\end{array}\right.
\qquad
 \Delta_-  =  \left\{\begin{array}{l}
           -8.08\qquad\text{our} \\
           -8.73 \qquad\text{Burgi}
\end{array}\right.
\]

The results for the polarizabilities evaluated with the central values 
for the LECs in Eqs.~(\ref{eq:LECsp4})-(\ref{eq:LECsp6}) are shown 
in Table~\ref{tab:polar1}.
\begin{table}[ph]
\tbl{Central values of polarizabilities in units of $10^{-4}\,{\rm fm}^3$. }
{\begin{tabular}{c||c|c}\toprule
    & to one loop & to two-loops 
\\[1.1ex] 
\hline
$(\alpha-\beta)_{\pi^+}$ \qquad & \qquad $ 6.0 $\qquad & \qquad $ 5.7  $
\\[1.1ex] 
$ (\alpha+\beta)_{\pi^+} $ \qquad & \qquad $ 0    $  \qquad & \qquad $ 0.16 $
\\ 
\hline
\end{tabular}
\label{tab:polar1}}
\end{table}
The uncertainty in the prediction for the polarizability has two sources.
First, the low-energy constants are not known precisely. Second,
we are dealing here  with an expansion in powers of the momenta and of the
quark masses up to and including terms of order~$p^6$. 
The discussion of estimating  uncertainties may be found in our
paper \cite{Gasser:2006qa}. 
It was shown that the value for the dipole polarizability 
$(\alpha-\beta)_{\pi^\pm}$ is rather reliable - there is no sign of any large, 
uncontrolled  correction to the two-loop result. 
The maximum deviation $1.0$ from the central value $5.7$ has been used 
as the final theoretical uncertainty for the dipole polarizability: 
\bea\label{eq:polarfinal}
(\alpha-\beta)_{\pi^\pm}= (5.7\pm 1.0)\times 10^{-4}\,{\rm fm}^3\,.
\eea
The chiral expansion for the combination $(\alpha+\beta)_{\pi^\pm}$
starts out at order $p^6$ so we have determined only its leading order term.

\section{Experimental information}

There are three types of experiments aiming to measure
the pion polarizabilities:

\begin{itemize}
\item
The scattering of high energy pions off the Coulomb 
field of heavy nucleus using the Primakoff effect

\item Radiative pion photoproduction from the proton

\item Pion pair production in photon-photon collisions

\end{itemize}

Schematically, they are shown in Fig.~\ref{fig:EXPT}.
\begin{figure}[bp]
\centerline{\includegraphics[width=10cm]{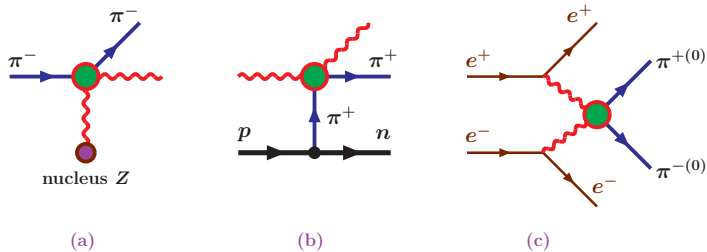}}
\caption{Schematic images of the available experiments.
\label{fig:EXPT}}
\end{figure}
%
%
The possibility to measure the pion polarizability via the Primakoff reaction
was proposed  in the early 1980s in \cite{Galperin:1980fa}.
The measurement of the pion-photon Compton scattering amplitude
by using the Primakoff effect was performed in an experiment
at Serpukhov \cite{Serpukhov} , but the small data sample led to 
only an imprecise value for the polarizability of 
$\alpha_\pi = ( 6.8\pm 1.4_{\rm stat} \pm 1.2_{\rm syst})\times 10^{-4}\, {\rm fm}^3$.
Low statistics made it difficult to evaluate the systematic uncertainty.

COMPASS has now achieved a modern Primakoff experiment, using a 190 GeV
pion beam from the Super Proton Synchrotron at CERN directed at a nickel
target. It is important that COMPASS was also able to use a muon, 
which is point-like particle, to calibrate the experiment. The Compton
$\pi^-\gamma\to \pi^-\gamma$ scattering is extracted from the reaction
$\pi^-\,Ni\to \pi^-\gamma\,Ni$ by selecting events from the Coulomb peak at 
small momentum transfer $Q^2<0.0015$~GeV$^2$. From the analysis of a sample of 
63,000 events, the collaboration obtained a value of the pion electric
polarizability of \cite{Adolph:2014kgj}
$\alpha_\pi = ( 2.0\pm 0.6_{\rm stat} \pm 0.7_{\rm syst})\times 10^{-4}\, {\rm fm}^3$
under assumption $(\alpha+\beta)_\pi=0$.
%

The cross section for the radiative pion photoproduction
$\gamma p\to\gamma\pi^+ n$ has been measured at the Lebedev Institute
\cite{Lebedev}. By using an extrapolation to the pion pole
in the unphysical region the value of the electric polarizability was obtained 
$\alpha_{\pi^+} = -\beta_{\pi^+}= (20 \pm 12)\times 10^{-4}\, {\rm fm}^3$.
Similar experiment was performed at the Mainz Microtron MAMI
\cite{MAMI} but the pion polarizability has been extracted by a comparison
of the data with the predictions of two different models yielding the value
$(\alpha-\beta)_{\pi^+} = (11.6 \pm 1.5_{\rm stat} \pm 3.0_{\rm syst} 
\pm 0.5_{\rm mod} )\times 10^{-4}\, {\rm fm}^3$.
%

Another possibility to obtain the value for the pion polarizability
is to extrapolate the data from  the pion pair production 
in photon-photon collisions $\gamma\gamma\to\pi\pi$ to
the region of the Compton scattering threshold by using
crossing symmetry and analyticity. Normally, the procedure involves
the construction of the dispersion relations with one or two subtractions.
The most recent analysis preformed in \cite{Fil'kov:2005ss} 
produced the value 
$(\alpha-\beta)_{\pi^\pm} = 13.0^{+2.6}_{-1.9} \times 10^{-4}\, {\rm fm}^3$ 
which is close to the MAMI data. There are plenty of previous studies
in this direction which give quite a broad region for the value
of the pion polarizability. The available experimental information
is shown in Table~\ref{tab:polar3}.

\begin{table}[ht]       
%
\tbl{Experimental information on $(\alpha-\beta)_{\pi^\pm}$,
 in units of $10^{-4}\,{\rm fm}^3$.
We indicate the reaction and data used.
In \protect\cite{DH93} ,
\protect\cite{Babusci} and \protect\cite{Adolph:2014kgj} 
$\alpha_\pi$ was determined, using as a constraint $\alpha_\pi = -\beta_\pi$. 
To obtain $(\alpha-\beta)_{\pi^\pm}$, we multiplied the results 
by a factor of 2.
}
{
\def\arraystretch{1.2}
\begin{tabular}{ll|c} \toprule
\multicolumn{2}{l|}{Experiments}    & $(\alpha-\beta)_{\pi^\pm}$  \\ 
\hline
 $\pi^-Z\to\gamma\pi^-Z$ & Serpukhov (1983)\,\cite{Serpukhov} & 
$15.6\pm 6.4_{\,\rm stat}\pm 4.4_{\,\rm syst}$ \\
\hline
$\gamma p\to \gamma \pi^+ n$ &
Lebedev Inst. (1984) \cite{Lebedev} & $ 40\pm 24 $ \\
\hline
\multicolumn{2}{l|}{D. Babusci {\em et al.} (1992)\, \cite{Babusci} } & \\
 $\gamma\gamma\to\pi^+\pi^-$ &   PLUTO \,\cite{PLUTO}  & 
 $38.2 \pm 9.6\pm 11.4$     \\
                             &    DM 1 \,\cite{DM1}  & $34.4\pm 9.2$  \\
                             &    DM 2 \,\cite{DM2} & $52.6\pm 14.8$ \\
                             & MARK II \,\cite{MARK-II} & $4.4\pm 3.2$\\
\hline
\multicolumn{2}{l|}{ J.F. Donoghue, B. Holstein (1993)\, \cite{DH93}  }  
& $5.4$ \\
$\gamma\gamma\to\pi^+\pi^-$ & MARK II\,\cite{MARK-II}  &       \\
\hline
\multicolumn{2}{l|}{ A. Kaloshin, V. Serebryakov (1994)\, \cite{KS94}} & 
$5.25\pm 0.95$\\ 
$\gamma\gamma\to\pi^+\pi^-$ & MARK II \,\cite{MARK-II}  & \\
&Crystal Ball Coll. \cite{crystalball}&\\
\hline
$\gamma p\to \gamma \pi^+ n$ & Mainz (2005) \cite{MAMI} & 
$11.6\pm 1.5_{\,\rm stat}\pm 3.0_{\,\rm syst} \pm 0.5_{\,\rm mod}$ \\
\hline
\multicolumn{2}{l|}{L. Fil'kov, V. Kashevarov (2005) 
\,\cite{Fil'kov:2005ss}}   & $13.0^{+2.6}_{-1.9}$  \\ 
 $\gamma\gamma\to\pi^+\pi^-$ & MARK II \,\cite{MARK-II}\,, & \\ 
\multicolumn{2}{l|}{ TPC/2$\gamma$ \cite{TPC}\,,\,   
 CELLO \,\cite{CELLO}\,, } & \\
\multicolumn{2}{c|}{VENUS\, \cite{VENUS}\,,   
 ALEPH \,\cite{ALEPH}\,,\, BELLE \cite{BELLE} }& \\
\hline
 $\pi^-Z\to\gamma\pi^-Z$ & COMPASS (2015)\,\cite{Adolph:2014kgj} & 
$ 4.0 \pm 1.2_{\,\rm stat}\pm 1.4_{\,\rm syst}$ \\
\hline
\end{tabular} \label{tab:polar3}}
\end{table}

\section*{Acknowledgments}

I would like to thank Juerg Gasser and Mikko Sainio for their collaboration
in which the results for pion polarizabilities at two-loop level 
accuracy were obtained. It has been a pleasure to discuss many points
of this subject with L.~V.~Fil'kov,  S.~B.~Gerasimov,  A.~G.~Olshevski, 
V.~A.~Petrunkin and M.~K.~Volkov. I am also grateful to 
Stanislav  Dubni\v{c}ka and Zuzana Dubni\v{c}kov\'{a} for inviting me 
to give a talk at the conference ``Hadron Structure'15'' . 



\end{document}